# Graphene-silicon device for visible and infrared photodetection

*Aniello Pelella[1,2], Alessandro Grillo[1,2], Enver Faella[1,2], Giuseppe Luongo[3], Mohammad Bagher Askari[4], and Antonio Di Bartolomeo[1,2]*

1 Department of Physics and Interdepartmental Centre NanoMates, University of Salerno, via Giovanni Paolo II, Fisciano, 84084, Italy

2 CNR-SPIN, via Giovanni Paolo II, Fisciano, 84084, Italy

3 IHP-Microelectronics, Im Technologie Park, Frankfurt Oder, Germany

4 Department of Physics, Faculty of Science, University of Guilan, P.O. Box: 41335-1914, Rasht, Iran

ABSTRACT

The fabrication of a graphene-silicon (Gr-Si) junction involves the formation of a parallel metal-insulator-semiconductor (MIS) structure, which is often disregarded but plays an important role in the optoelectronic properties of the device. In this work, the transfer of graphene onto a patterned n-type Si substrate, covered by $Si_3N_4$, produces a Gr-Si device in which the parallel MIS consists of a Gr-$Si_3N_4$-Si structure surrounding the Gr-Si junction. The Gr-Si device exhibits rectifying behavior with a rectification ratio up to $10^4$. The investigation of its temperature behavior is necessary to accurately estimate the Schottky barrier height at zero bias, $\varphi_{b0} = 0.24\ eV$, the effective Richardson's constant, $A^* = 7 \times 10^{-10}\ AK^{-2}cm^{-2}$, and the diode ideality factor. n=2.66 of the Gr-Si junction. The device is operated as a photodetector in both photocurrent and photovoltage mode in the visible and infrared (IR) spectral regions. A responsivity up to 350 mA/W and external quantum efficiency (EQE) up to 75% is achieved in the 500-1200 nm wavelength range. A decrease of responsivity to 0.4 mA/W and EQE to 0.03% is observed above 1200 nm, that is in the IR region beyond the silicon optical bandgap, in which photoexcitation is driven by graphene. Finally, a model based on two back-to-back diodes, one for the Gr-Si junction the other for the Gr-Si$_3$N$_4$-Si MIS structure, is proposed to explain the electrical behavior of the Gr-Si device.

INTRODUCTION

Silicon has been leading the development of the semiconductor technology holding a dominant position in the microelectronics field for decades. However, silicon as an optoelectronic material suffers from the short bandwidth and the large surface reflectivity that limit the responsivity and the application of silicon-based photodetectors to near-infrared bands. Near infrared (NIR) photodetection, particularly at 1550 nm, is crucial for a variety of applications, ranging from optical communications[1–3] to remote sensing[4–6].

Graphene offers a very attractive platform for advanced optoelectronic applications, due to its high conductivity, zero bandgap, low noise, flexibility, chemical stability and other extraordinary properties[7–12]. Owing to its semimetal behavior and tunable energy Fermi level, it enables new functionalities in traditional electronic and optoelectronic devices[13,14]. For instance, silicon can be joined to graphene to form a Schottky diode which is used as bias-controlled photodetector[14–18]. Graphene can replace the metal contact of a Schottky junction and make shallow junctions with enhanced photoresponse[19–21]. In reverse bias, photons with energy higher than the semiconductor band gap, absorbed in the semiconductor depletion layer, induce the formation of photocharges that are separated by the junction built-in field, originating a photocurrent. Moreover, photons with sub-bandgap energy can be adsorbed by graphene and inject electrons over the Gr-Si barrier, leading to a charge flow from the graphene to the semiconductor[21,22]. Furthermore, the Gr-Si junction is a basic element for novel electronic devices for the integration of graphene into the existing semiconductor technology[23–25].

A Gr-Si junction is fabricated using a Si substrate covered by an $SiO_2$ dielectric layer, typically 100-300 nm thick. The etching of a window in the $SiO_2$ cap layer exposes the Si area for the formation of the Gr-Si junction. The transferred graphene covers the bare Si area and encroaches upon the oxide layer for the formation of contacts with metal leads. Such an encroachment of graphene over $SiO_2$ originates a MIS structure, namely a Gr-$SiO_2$-Si structure, that is in parallel with the Gr-Si junction. It has been shown that the MIS structure affects the current-voltage (IV) and capacitance-voltage characteristics of the junction[26] and enhances its photodetection capability[27–29]. Despite its important role, the parallel MIS structure is often neglected in applications using the Gr-Si junctions.

In this work, we transferred graphene monolayers, produced by chemical vapor deposition (CVD), onto a n-Si wafer covered by a patterned $Si_3N_4$ layer. In such a way, we formed Gr-Si junctions in parallel with Gr-$Si_3N_4$-Si MIS structures. To clarify the role of the parallel MIS, we investigated the optical and electronic properties of the combined device, referred to as the Gr-Si device. We evaluated transport parameters such as the Gr-Si Schottky barrier height (SBH) and the ideality factor and we studied the responsivity of the device in the visible and near-infrared region demonstrating a promising photodetector. Finally, we proposed a model based on two back-to-back diodes to explain the experimental findings.

MATERIALS AND METHODS

Samples were prepared on doped n-Si (100) wafers with resistivity of ~10 Ωcm, corresponding to a phosphorus dopant density of $\sim 4.5 \times 10^{14}\ cm^{-3}$. A $15\ nm$ thick silicon nitride ($Si_3N_4$) was deposited by CVD. Then, a $3x3\ mm^2$ trench was patterned by lithography and wet etching of the silicon nitride. The trench area was further cleaned by hydrofluoric acid immediately before the Gr deposition to prevent or limit the formation of native oxide.

A $\sim 5 \times 7\ mm^2$ Gr sheets was transferred onto the Si substrates by a wet method (the details are reported elsewhere[30]) to cover the Si trench while extending over the surrounding $Si_3N_4$ layer, thus acting both as the anode of the Gr-Si junction and the top (gate) electrode of the Gr-$Si_3N_4$-Si MIS structure. Ohmic metal contacts to graphene were realized by Ag paste. Likewise, Ag paste was spread on the scratched backside of the Si-substrate to guarantee an ohmic back contact.

A device schematic is reported in Figure 1(a). Furthermore, a Raman line-scan (Figure1(b)) provides a clear evidence of a high-quality monolayer Gr layer, confirmed by the high 2D/G intensity peak ratio and the negligible defect-related D-peak ($\sim 1350\ cm^{-1}$).

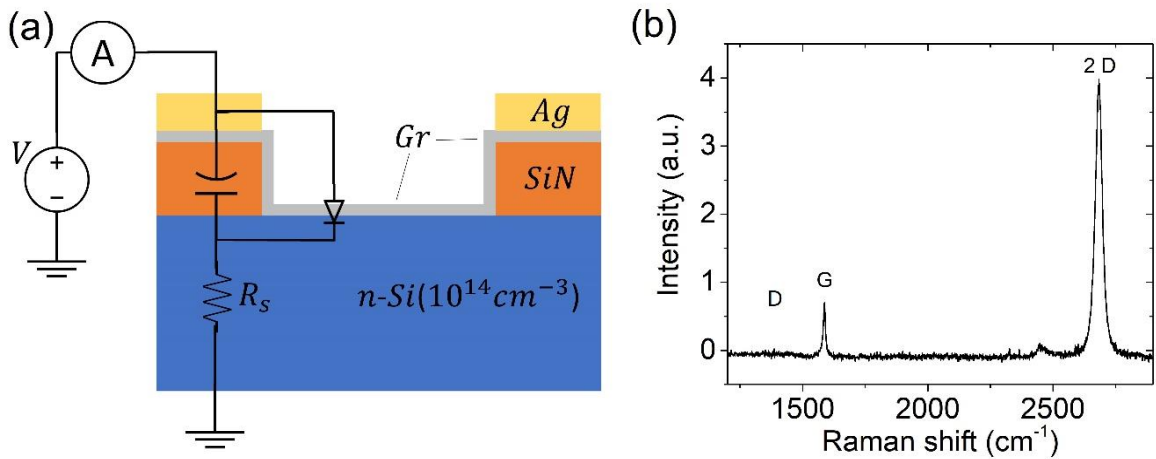

**Figure 1.** (a) Device schematic showing a Gr-Si junction modelled by a diode in parallel with a MIS structure, here modeled as a capacitor. (b) Raman spectroscopy confirming high quality monolayer graphene.

RESULTS AND DISCUSSION

Figure 2(a) shows the semi-logarithmic plot of the IV characteristic of the Gr-Si device in dark at $300\ K$ ad atmospheric pressure. The device exhibits a rectifying behavior, with rectification ratio $\sim 10^4$ at $V = \pm 2.5\ V$ and shape which suggests the use of the diode equation for the current:

$$I = I_0 \left[\exp\left(\frac{qV}{nkT}\right) - 1\right] \quad (1)$$

where $I_0$ is the reverse saturation current, $q$ is the electron charge, $n$ the ideality factor, $k$ the Boltzmann constant and $T$ the temperature. The ideality factor considers deviations from pure thermoelectric transport ($n = 1$) which can take place in a Schottky diode. The presence of defects or unwanted insulating layers can cause Schottky barrier inhomogeneities and increase the ideality factor. The green dashed line in Figure 2(a) represents the fit to the experimental data of Eq. (1), with $I_0$ and $n$ as fitting parameters, obtained in the forward region $0 < V < 0.4\ V$. Indeed, in this region, any series resistance can be neglected, and the best fit is obtained with $n = 2.02$ and $I_0 = 1.69 \cdot 10^{-9}\ A$. Using the estimated value of $I_0$ and referring to the thermionic theory:

$$I_0 = SA^*T^2 \exp\left(-\frac{\varphi_{b0}}{kT}\right) \rightarrow \varphi_{b0} = kT\ln\left(\frac{SA^*T^2}{I_0}\right) = 0.88\ eV \quad (2)$$

where $S = 0.09\ cm^2$ is the area of the junction, $A^* = 112\ cm^{-2}K^{-2}$ is the Richardson constant for n-Si and $\varphi_{b0}$ is the SBH at zero voltage. This estimation is affected by an error which is related to the arbitrary choice of the forward region for the fit and might contain a huge systematic error related to the assumed values of S and $A^*$ (see following).

A more realistic model includes a series resistance $R_s$. We then follow Cheung's method[31] for evaluating the diode parameters. According to Cheung's method, Eq. (1) with a series resistance becomes:

$$I = I_0 \left[\exp\left(\frac{q(V - R_s I)}{nkT}\right) - 1\right] \quad (3)$$

which, for $V - R_s I \gg nkT/q$, provides:

$$\frac{dV}{d\ln I} = R_s I + \frac{nkT}{q} \quad (4)$$

From the fit of Eq. (3), we can evaluate $R_s$ and $n$, which can be used to estimate $\varphi_{b0}$, defining the following function:

$$H(I) = V - \frac{nkT}{q}\ln\left(\frac{I}{SA^*T^2}\right) = R_s I + \frac{n}{q}\varphi_{b0} \tag{5}$$

Figure 2(b) shows the plots of $\frac{dV}{d\ln I}$ and $H(I)$ a function of $I$ from which we extract the following parameters: $n = 2.8$, $Rs = 85\ k\Omega$ and $\varphi_{b0} = 0.87\ eV$. Using the so estimated $I_0$, $R_s$, and $n$ in Eq. (3), we obtain the magenta dotted curve shown in Figure 2(a). Such a curve provides a good fit to the forward current, till the reach of the flat band condition ($V - R_s I \simeq 0$, which occurs at $V \simeq 0.59V$), where the diode equation does not apply anymore.

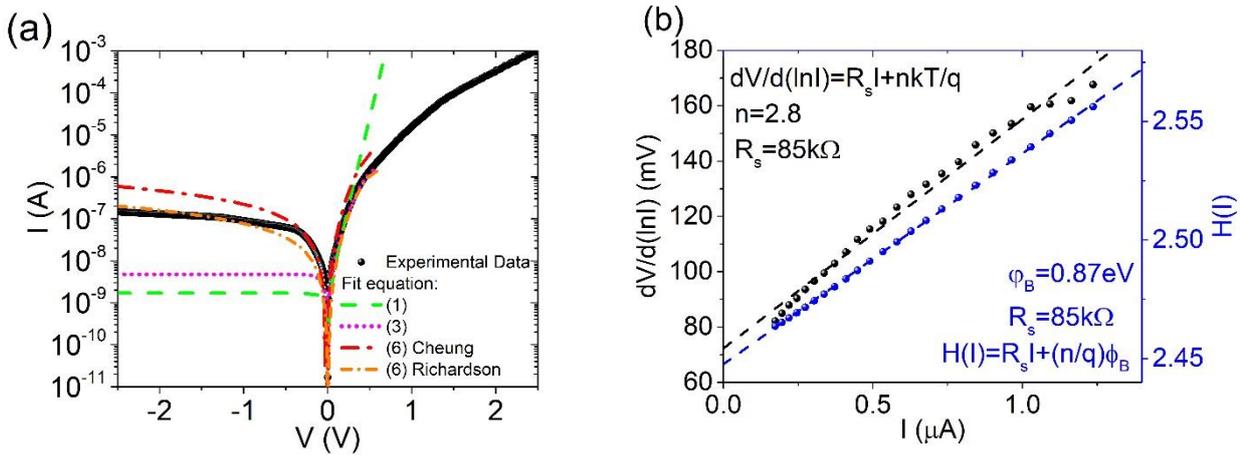

**Figure 2.** (a) Measured IV characteristic of the device in dark (black line). The green and magenta lines represent the fit using respectively Eq. (1) and (3). Red and orange lines represent the fit using Eq. (6) with Cheung and Richardson's parameters, respectively. (b) Cheung's method plots for evaluation of Schottky barrier height, ideality factor and series resistance.

As all fittings fail in the reverse region, we further improve the model by considering a parallel resistance Rp[27]:

$$I = \frac{R_p}{R_s + R_p}\left\{I_0\left[\exp\left(\frac{q(V - R_s I)}{nkT}\right) - 1\right] - \frac{V}{R_p}\right\} \tag{6}$$

Eq. (6) provides a good fit both in reverse and forward bias, as shown by the red dash-dot line in Figure 2(a), when $R_p = 4\ M\Omega$ and $I_0 = 2.3 \times 10^{-8}\ A$.

Figure 3(a) and 3(b) display the IV characteristics of the Gr-Si device at different temperatures, from $T = 400K$ to $T = 220K$, in dark and under illumination by a supercontinuum laser ($65\ mW/cm^2$, $\lambda = 500nm$), respectively. The plots show that lowering the temperature suppresses both the forward and the reverse currents, as predicted by the thermionic theory (Eq. (1) and Eq. (2)). Noteworthy, a substantial difference between the current in dark and under illuminations is observed only at temperatures below $320\ K$ and in reverse bias. Light does not appreciably change the forward current and, at higher temperatures, the reverse current. At low temperatures, the IV curves present a sudden curvature changes (one or more kinks) in reverse bias probably caused by the parallel MIS capacitor, as explained next.

Figure 3(c) shows the rectification ratio at $V = \pm 2.5\ V$. In dark, the rectification ratio increases with the decreasing temperature, indicating that the Schottky barrier becomes more efficient in suppressing the electron flow from graphene to silicon (reverse current) when the temperature is lowered. Under illumination, the rectification ratio overlaps that measured in dark for temperatures higher than $320\ K$, i.e., when thermal generation overcomes photogeneration. Below $320\ K$, the rectification ratio decreases with the decreasing temperature because the reverse current becomes more and more dominated by the photogeneration. This result indicates that the suppressed dark (reverse) current at low temperature is favourable to photodetection. Indeed, at low temperature, the reverse current is substantially enhanced by the electron-hole photogeneration.

From the Cheung's method, we obtained value of $n$, $R_s$, and $\varphi_{b0}$ as function of the temperature, respectively reported in Figure 3(d), 3(e) and 3(f), in dark (black dot-lines) and under light (red dot-lines). The temperature dependence of the ideality factor in Figure 3(d) shows a descending behaviour for increasing temperature. This feature indicates that deviations from the ideal thermionic behaviour of the diode are more effective at lower temperatures when thermionic emission is suppressed and tunnelling and diffusion might become comparatively relevant. The dependence on the temperature of the series resistance is typical of a semiconductor and is probably dominated by the silicon substrate.

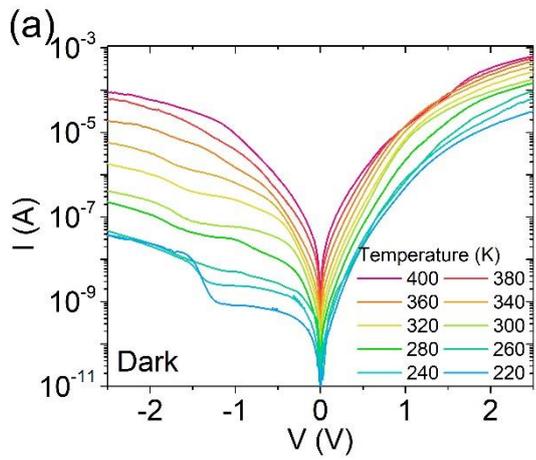
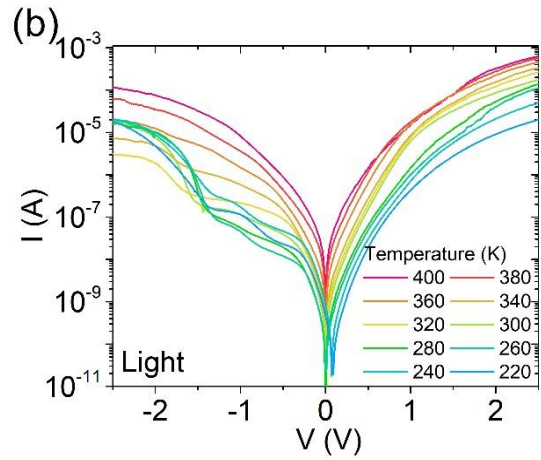
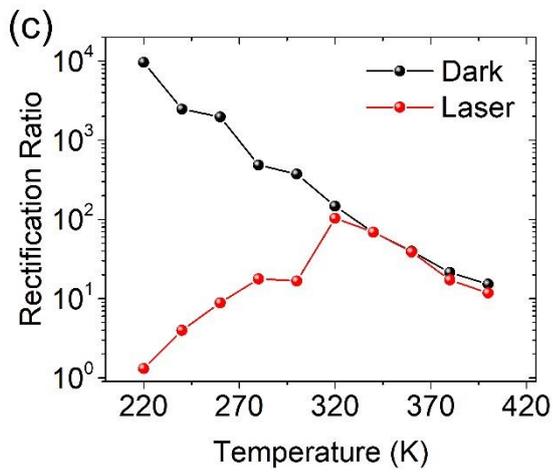
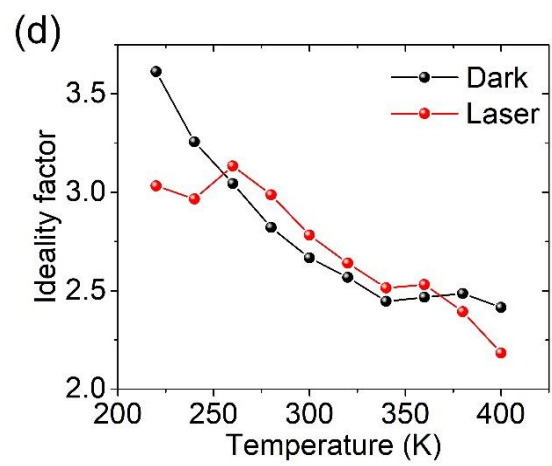
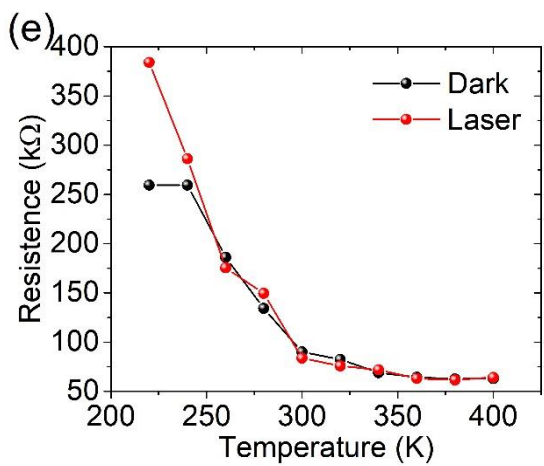
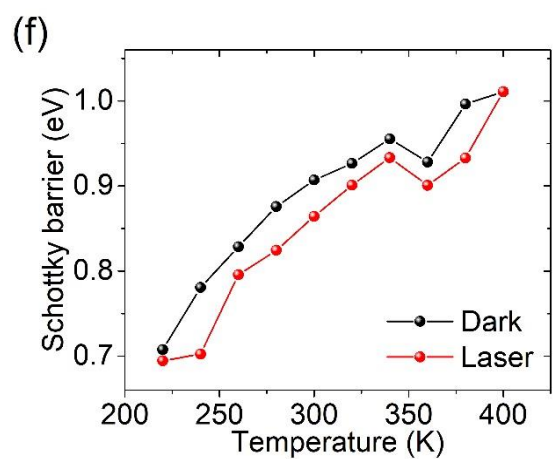

**Figure 3.** IV characteristics versus the temperature, ranging from $400\ K$ to $220\ K$ in (a) dark and (b) light. (c) Rectification ratio at $V = \pm 2.5\ V$, (d) ideality factor, (e) series resistance, (f) Schottky barrier height versus temperature, estimated using the Cheung's method.

Finally, Figure 3(f) shows that the SBH decreases with temperature. This is a well-known effect when there is barrier inhomogeneity[32–34]. The reduced thermal energy makes carrier cross the barrier mainly in the position where the SBH is lower, thus resulting in a reduced SBH. This effect is less pronounced under illumination where photogenerated charges have enough energy to cross the barrier everywhere.

The obtained zero bias SBH, $\varphi_{b0} = 0.87\ eV$, although consistent with some previous works[35,36], exceeds the prediction of the Schottky-Mott model, $\varphi_{b0} = \Phi_g - \chi_{Si} = 0.5\ eV$ ($\Phi_g = 4.54$ eV is the commonly used work function of graphene and $\chi_{Si} = 4.05\ eV$ is the electronic affinity of Si) and should result in lower reverse current and higher rectification ratio than observed. We note that a large discrepancy can be found in the literature for the Gr-Si SBH (from 0.2 to 0.9 eV[27,35–41]) estimated from electrical transport and that complementary techniques such as to X-ray photoemission spectroscopy lead to much lower barrier height[39]. Furthermore, several experimental studies of the Gr-Si Schottky junction have indicated that an effective Richardson constant, order of magnitude lower than $112\ cm^{-2}K^{-2}$, is needed to account for the experimental data[25,26,32,33,42,43]. Therefore, to make an estimation of the SBH independent of the effective Richardson constant $A^*$ and to avoid relying on a single IV characteristic, we extracted the SBH and, as by-product, the effective Richardson constant from the IV characteristics measured at different temperature (Figures 3(a) and 3(b)). According to Eq. (2), $\varphi_{b0}$ and $A^*$ can be obtained from the slope and intercept of $\ln(I_0/T^2)$ versus $\frac{q}{kT}$ (Richardson's plot in Figures 4(c) and 4(d)). $I_0$ is the extrapolated current at $V = 0\ V$ obtained from the linear fittings of the forward current shown in Figures 4(a) and 4(b). The estimated $A^* \approx 7 \times 10^{-10} Acm^{-2}K^{-2}$ is significantly lower than the theoretical one ($112\ A\ cm^{-2}K^{-2}$) used for previous calculations, and the obtained SBH reduces to $\varphi_{b0} = 0.24\ eV$, consistent with the modest rectification ratio. Although the origin of a lower effective Richardson constant is still under debate[16,22], the inadvertent presence of a native oxide layer at Gr-Si interface[27], the massless Dirac fermion nature of carriers in graphene[15,16] or a Landauer transport mechanism[42] have been invoked as an explanation. We point out that the SBH does not change when measured in dark or under illumination. The $I_0$ obtained by Eq. (2) with $\varphi_{b0}$ and $A^*$ from Richardson's plot, inserted in Eq. (6), provide the best fit to the data, as shown by orange dash-dot line in Figure 2(a) ( the other fitting parameters are $R_p = 12\ M\Omega$, $I_0 = 5.85 \times 10^{-9}\ A$). The excellent agreement between experimental data and model is taken as a confirmation of the higher accuracy of the Richardson plot-based method.

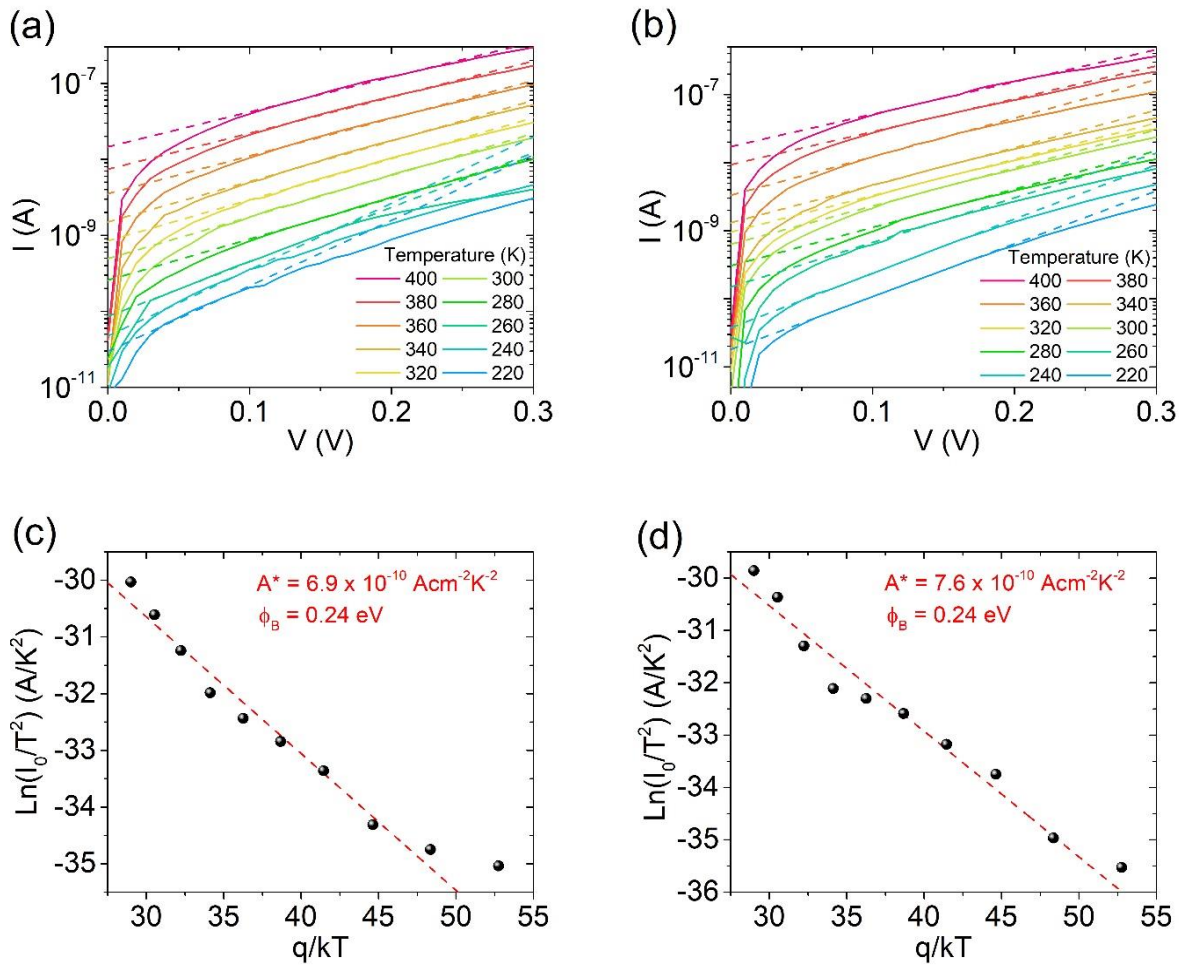

**Figure 4.** Linear fits used to extract $I_0$ at $V = 0\ V$ (a) in dark and (b) under illumination. Richardson's plots obtained from IV measurements in (c) dark and (d) under illumination.

To further investigate the physical mechanisms of charge transport, we tested the optical response of the device. Figure 5(a) shows the IV characteristics in dark and under illumination by a supercontinuum white laser. Tuning the emission power of the laser from 0 to $180\ mW$ (corresponding to $0-1800\ mW/cm^2$ light intensity), we obtained the photoresponse of the device, defined as $(I_{light}-I_{dark})/I_{dark}$ at $V=-2.5\ V$ shown in Figure 5(b). The data follow an exponential law with the e-folding factor $\tau=56\ mW$, which ensures that a maximum light effect is reached above $\sim 170\ mW$. Accordingly, the laser was set to the maximum power of $180\ mW$ for further tests of the Gr-Si device in photocurrent and photovoltage mode.

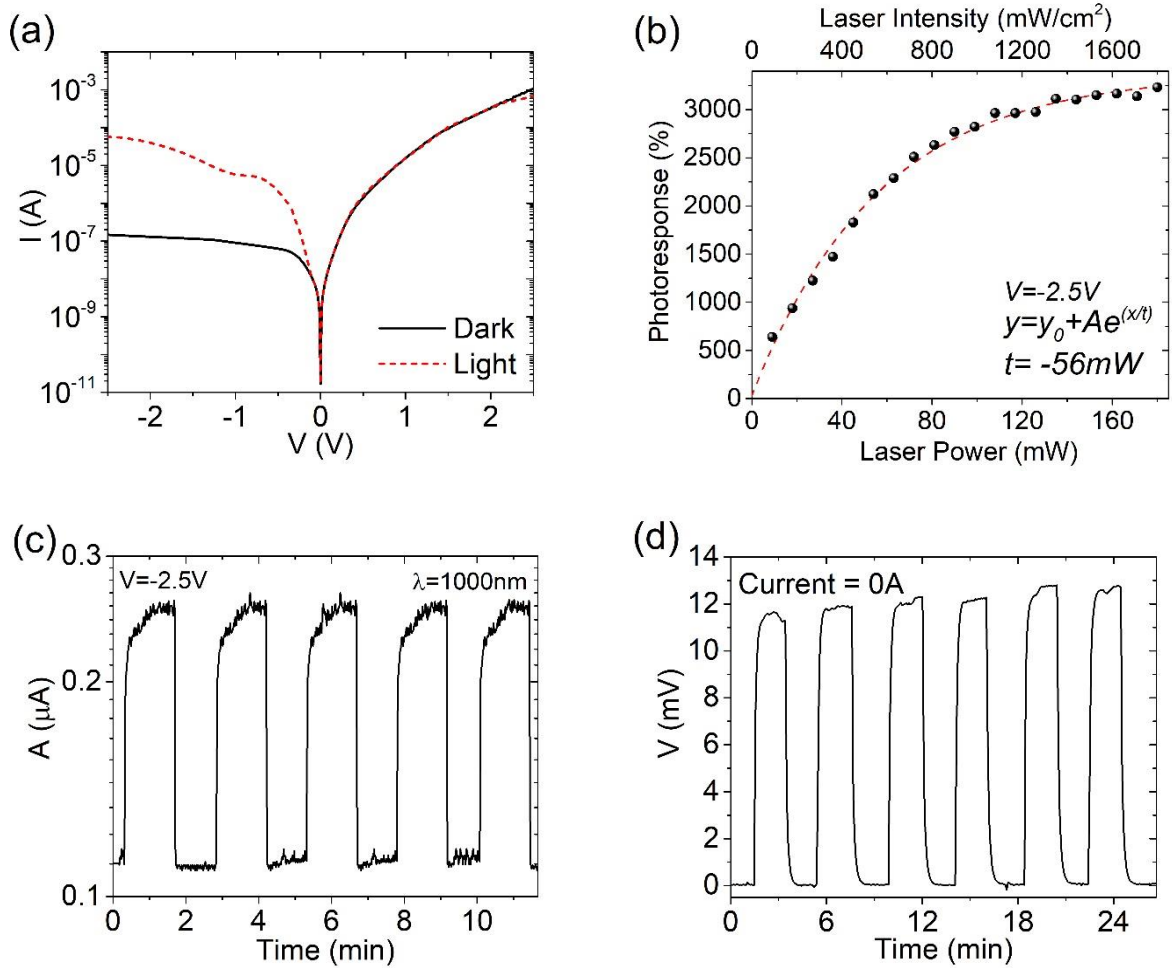

**Figure 5.** a) IV characteristics in dark and with incident white laser. b) Photoresponse as a function of the laser intensity from 0 to $1800\ mW/cm^2$. c) Photocurrent when the photodetector is operated in photocurrent mode at $V=-2.5\ V$ and d) photovoltage mode at $I=0\ A$) under laser beam with $1000\ nm$ wavelength and $180\ mW$ power.

Setting the voltage to $V = -2.5\,V$, we performed a series of measurements exposing it to the laser beam with wavelength emission of $\lambda = 1000\,nm$. The device reacts to the laser with fast and repeatable photocurrent (Figure 5(c)). Furthermore, at zero current, we also observed a photovoltaic effect as reported in Figure 5(d)) where a voltage around 12 mV appears at the electrode in response to a laser pulse.

To complete the optoelectronic characterization of the device, we investigated the spectral response in the 500-2000 $nm$ wavelength range by sampling the spectrum of the supercontinuum laser in intervals of 50 $nm$ and 20 $nm$ bandwidth. Figure 6(a) reports the responsivity of the device, defined as the ratio of photocurrent to the incident power, $R = (I_{light} - I_{dark})/P_{inc}(\lambda)$ along with external quantum efficiency $EQE = \frac{(I_{light}-I_{dark})}{q}\frac{hc}{P_{inc}(\lambda)\lambda} = \frac{R}{\lambda(nm)} \times 1240$ ($\lambda$ is the wavelength, h is the Planck constant, and c is the speed of light). It shows an EQE around 75% for $\lambda < 1100\,nm$, i.e. when photoconversion occurs main in Si, followed by a sudden drop to 0.03% when the energy of the incident light is below the bandgap of Si. For $\lambda > 1100\,nm$, photoexcitation occurs mainly in graphene and the EQE, as reported also elsewhere[10,44]. We highlight that the obtained external quantum efficiency is among the highest reported in the literature over the investigated spectral range[45–48].

As an additional figure of merit, Figure 6(b) shows Noise Equivalent Power ($NEP = \frac{\sqrt{2qI_{dark}}}{R}$) that indicates the minimum detectable power. Obviously, the higher quantum efficiency corresponds to lower detection power.

Figures 6(c) and 6(d) show the IV characteristics under light at different wavelengths. It can be observed that a kink forms at about $V = -1.2\,V$ in the reverse curves. The photocurrent at given illumination and wavelength is reaches a plateau at high reverse bias, when it is limited by photocarrier generation rate.

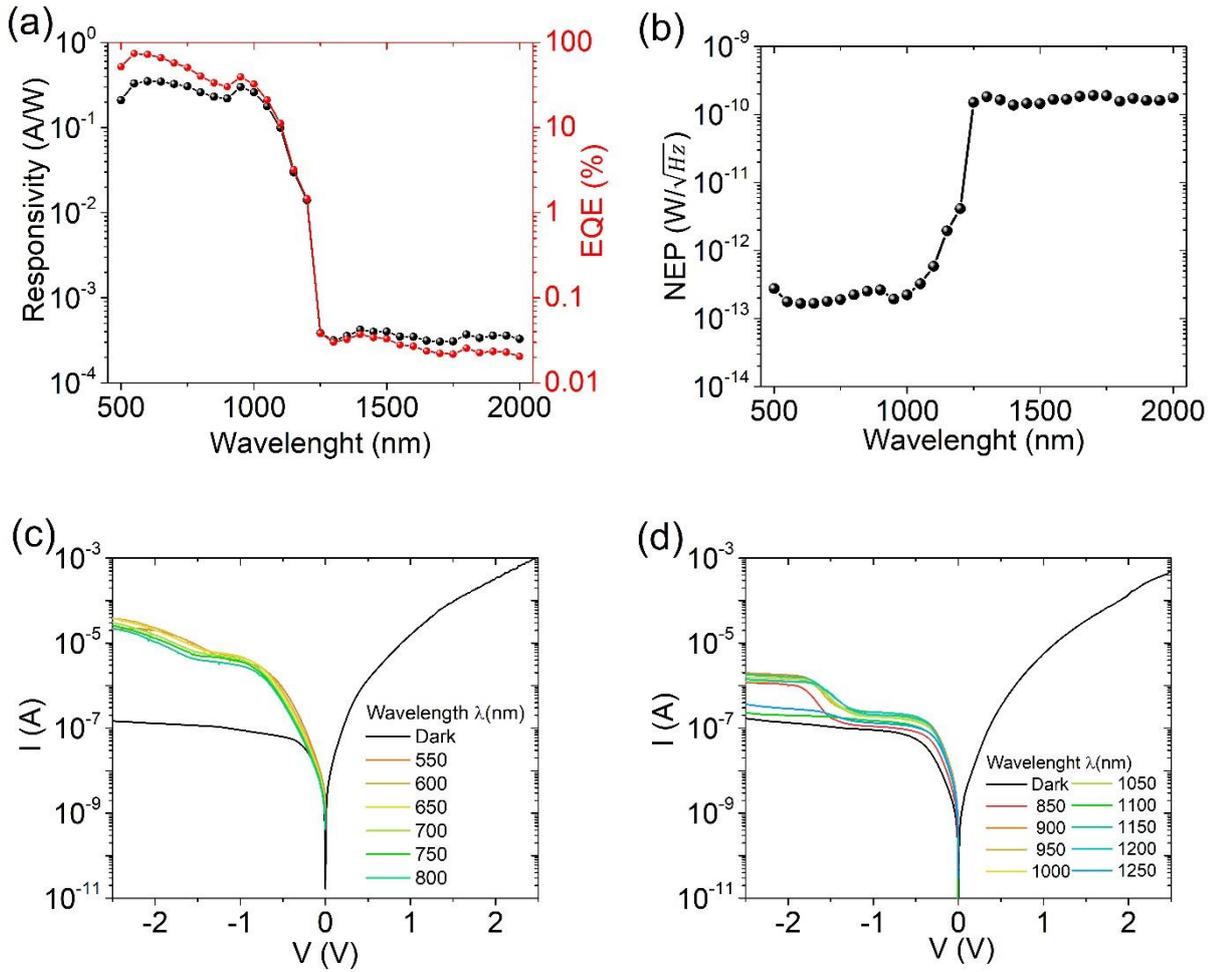

**Figure 6.** (a) Responsivity and EQE of the device in the visible and IR spectral region. (b) NEP and (c) IV characteristics in dark and under light at different wavelengths in the (c) visible and (d) near spectral IR regions.

The observed optoelectronic behavior of the Gr-Si device can be understood by considering the parallel Gr-Si$_3$N$_4$-n-Si structure, which in forward behaves like a MIS capacitor charged by electron on the Si side. Such electron can diffuse to the Gr-n-Si junction and contribute to the forward current. In reverse bias, the negative voltage attracts holes at the Si-Si$_3$N$_4$ interface. As the holes accumulate, the Si undergoes an inversion and becomes p-type. When the voltage is high enough to enable tunneling through the gate oxide a p-type diode is formed in the MIS region. This means that, in reverse bias, the device behaves as two parallel and opposite diodes, a reverse Schottky diode due to the Gr-Si junction and a forward MIS diode formed by the Gr-Si$_3$N$_4$-n-Si structure. This parallel configuration explains the kinks at about $V = -1.2\, V$. Indeed, for $-1.2\, V < V < 0\, V$, holes accumulated at the interface Si-Si$_3$N$_4$ can only diffuse towards the Gr-Si junction and contribute to its reverse current (Figure 7(a)) originating the high leakage of $\sim 10^{-7}\ A$. For $V < -1.2\ V$, the electric field enables Fowler-Nordheim tunneling[49,50] through the Si$_3$N$_4$ layer (Figure 7(b)), resulting in an increase of current, which generates the aforementioned kinks. A current plateau is reached at high reverse voltage because of the thermal or photo generation limited rate in Si.

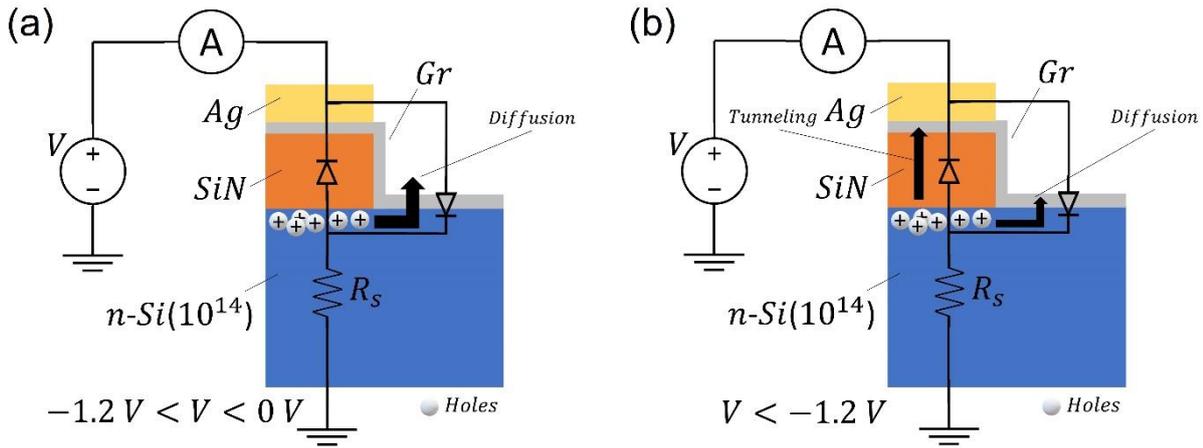

**Figure 7.** Schematic model of the Gr-Si device and charge carrier transport in reverse bias, for (a) $-1.2\ V < V < 0\ V$ and (b) $V < -1.2\ V$.

CONCLUSIONS

We have studied a Schottky Gr-Si junction in parallel with a Gr-Si$_3$N$_4$-Si structure, forming a composite Gr-Si device with high responsivity and external quantum efficiency in the visible and near infrared region. The device can be operated both in photocurrent and photovoltage mode. We have evaluated the relevant parameters of the junction and shown that only a detailed IV-T analysis leads to realistic estimation of the Gr-Si Schottky barrier parameters. We have detected the appearance of a kink in the reverse current, and we have proposed a model based on two parallel back-to-back diodes to explain it. We have clarified how the parallel Gr-Si$_3$N$_4$-Si structure introduced in the fabrication of the device contributes to the optoelectronic properties the Gr/Si heterostructure.

AUTHOR INFORMATION

**Corresponding Author**

**Antonio Di Bartolomeo** - Department of Physics and Interdepartmental Centre NanoMates, University of Salerno, via Giovanni Paolo II, Fisciano, 84084, Italy; email: adibartolomeo@unisa.it

**Aniello Pelella** - Department of Physics and Interdepartmental Centre NanoMates, University of Salerno, via Giovanni Paolo II, Fisciano, 84084, Italy; email: apelella@unisa.it

**Author Contributions**

The manuscript was written through contributions of all authors. All authors have given approval to the final version of the manuscript.

ACKNOWLEDGEMENT

A.D.B. thanks the University of Salerno, Salerno, Italy for the grants ORSA200207 and ORSA195727.

REFERENCES

(1) Agrawal, G. P. *Fiber-Optic Communication Systems*; John Wiley & Sons, 2012; Vol. 222.
(2) Majumdar, A. K.; Ricklin, J. C. *Free-Space Laser Communications: Principles and Advances*; Springer New York: New York, NY, 2008. https://doi.org/10.1007/978-0-387-28677-8.
(3) Ijaz, M.; Ghassemlooy, Z.; Rajbhandari, S.; Minh, H. L.; Perez, J.; Gholami, A. Comparison of 830 Nm and 1550 Nm Based Free Space Optical Communications Link under Controlled Fog Conditions. In *2012 8th International Symposium on Communication Systems, Networks Digital Signal Processing (CSNDSP)*; 2012; pp 1–5. https://doi.org/10.1109/CSNDSP.2012.6292739.
(4) Blackett, M. An Overview of Infrared Remote Sensing of Volcanic Activity. *J. Imaging* **2017**, *3* (2), 13. https://doi.org/10.3390/jimaging3020013.
(5) Goddijn-Murphy, L.; Williamson, B. On Thermal Infrared Remote Sensing of Plastic Pollution in Natural Waters. *Remote Sensing* **2019**, *11* (18), 2159. https://doi.org/10.3390/rs11182159.
(6) Hua, L.; Shao, G. The Progress of Operational Forest Fire Monitoring with Infrared Remote Sensing. *J. For. Res.* **2017**, *28* (2), 215–229. https://doi.org/10.1007/s11676-016-0361-8.
(7) Geim, A. K. Graphene: Status and Prospects. *Science* **2009**, *324* (5934), 1530–1534. https://doi.org/10.1126/science.1158877.
(8) Giubileo, F.; Martucciello, N.; Di Bartolomeo, A. Focus on Graphene and Related Materials. *Nanotechnology* **2017**, *28* (41), 410201. https://doi.org/10.1088/1361-6528/aa848d.
(9) Wang, J.; Mu, X.; Sun, M.; Mu, T. Optoelectronic Properties and Applications of Graphene-Based Hybrid Nanomaterials and van Der Waals Heterostructures. *Applied Materials Today* **2019**, *16*, 1–20. https://doi.org/10.1016/j.apmt.2019.03.006.

(10) Luo, F.; Zhu, M.; tan, Y.; Sun, H.; Luo, W.; Peng, G.; Zhu, Z.; Zhang, X.-A.; Qin, S. High Responsivity Graphene Photodetectors from Visible to Near-Infrared by Photogating Effect. *AIP Advances* **2018**, *8* (11), 115106. https://doi.org/10.1063/1.5054760.

(11) Bartolomeo, A. D.; Giubileo, F.; Santandrea, S.; Romeo, F.; Citro, R.; Schroeder, T.; Lupina, G. Charge Transfer and Partial Pinning at the Contacts as the Origin of a Double Dip in the Transfer Characteristics of Graphene-Based Field-Effect Transistors. *Nanotechnology* **2011**, *22* (27), 275702. https://doi.org/10.1088/0957-4484/22/27/275702.

(12) Urban, F.; Lupina, G.; Grillo, A.; Martucciello, N.; Di Bartolomeo, A. Contact Resistance and Mobility in Back-Gate Graphene Transistors. *Nano Express* **2020**, *1* (1), 010001. https://doi.org/10.1088/2632-959X/ab7055.

(13) Di Bartolomeo, A.; Santandrea, S.; Giubileo, F.; Romeo, F.; Petrosino, M.; Citro, R.; Barbara, P.; Lupina, G.; Schroeder, T.; Rubino, A. Effect of Back-Gate on Contact Resistance and on Channel Conductance in Graphene-Based Field-Effect Transistors. *Diamond and Related Materials* **2013**, *38*, 19–23. https://doi.org/10.1016/j.diamond.2013.06.002.

(14) Di Bartolomeo, A. Graphene Schottky Diodes: An Experimental Review of the Rectifying Graphene/Semiconductor Heterojunction. *Physics Reports* **2016**, *606*, 1–58. https://doi.org/10.1016/j.physrep.2015.10.003.

(15) Liang, S.-J.; Ang, L. K. Electron Thermionic Emission from Graphene and a Thermionic Energy Converter. *Phys. Rev. Applied* **2015**, *3* (1), 014002. https://doi.org/10.1103/PhysRevApplied.3.014002.

(16) Liang, S.; Hu, W.; Bartolomeo, A. D.; Adam, S.; Ang, L. K. A Modified Schottky Model for Graphene-Semiconductor (3D/2D) Contact: A Combined Theoretical and Experimental Study. In *2016 IEEE International Electron Devices Meeting (IEDM)*; 2016; p 14.4.1-14.4.4. https://doi.org/10.1109/IEDM.2016.7838416.

(17) Casalino, M.; Sassi, U.; Goykhman, I.; Eiden, A.; Lidorikis, E.; Milana, S.; De Fazio, D.; Tomarchio, F.; Iodice, M.; Coppola, G.; Ferrari, A. C. Vertically Illuminated, Resonant Cavity Enhanced, Graphene–Silicon Schottky Photodetectors. *ACS Nano* **2017**, *11* (11), 10955–10963. https://doi.org/10.1021/acsnano.7b04792.

(18) Alvarado Chavarin, C.; Strobel, C.; Kitzmann, J.; Di Bartolomeo, A.; Lukosius, M.; Albert, M.; Bartha, J.; Wenger, C. Current Modulation of a Heterojunction Structure by an Ultra-Thin Graphene Base Electrode. *Materials* **2018**, *11* (3), 345. https://doi.org/10.3390/ma11030345.

(19) Luongo, G.; Giubileo, F.; Iemmo, L.; Di Bartolomeo, A. The Role of the Substrate in Graphene/Silicon Photodiodes. *J. Phys.: Conf. Ser.* **2018**, *956*, 012019. https://doi.org/10.1088/1742-6596/956/1/012019.

(20) Casalino, M.; Crisci, T.; Moretti, L.; Gioffre, M.; Iodice, M.; Coppola, G.; Maccagnani, P.; Rizzoli, R.; Bonafe, F.; Summonte, C.; Morandi, V. Silicon Meet Graphene for a New Family of Near-Infrared Resonant Cavity Enhanced Photodetectors. In *2020 22nd International Conference on Transparent Optical Networks (ICTON)*; IEEE: Bari, Italy, 2020; pp 1–4. https://doi.org/10.1109/ICTON51198.2020.9203222.

(21) Di Bartolomeo, A.; Luongo, G.; Iemmo, L.; Urban, F.; Giubileo, F. Graphene–Silicon Schottky Diodes for Photodetection. *IEEE Trans. Nanotechnology* **2018**, *17* (6), 1133–1137. https://doi.org/10.1109/TNANO.2018.2853798.

(22) Varonides, A. Combined Thermionic and Field Emission Reverse Current for Ideal Graphene/n-Si Schottky Contacts in a Modified Landauer Formalism. *physica status solidi c* **2016**, *13* (10–12), 1040–1044. https://doi.org/10.1002/pssc.201600096.

(23) Wu, H.-Q.; Linghu, C.-Y.; Lu, H.-M.; Qian, H. Graphene Applications in Electronic and Optoelectronic Devices and Circuits. *Chinese Physics B* **2013**, *22*, 098106. https://doi.org/10.1088/1674-1056/22/9/098106.

(24) Hong, S. K.; Kim, C. S.; Hwang, W. S.; Cho, B. J. Hybrid Integration of Graphene Analog and Silicon Complementary Metal–Oxide–Semiconductor Digital Circuits. *ACS Nano* **2016**, *10* (7), 7142–7146. https://doi.org/10.1021/acsnano.6b03382.

(25) Luongo, G.; Grillo, A.; Giubileo, F.; Iemmo, L.; Lukosius, M.; Alvarado Chavarin, C.; Wenger, C.; Di Bartolomeo, A. Graphene Schottky Junction on Pillar Patterned Silicon Substrate. *Nanomaterials* **2019**, *9* (5), 659. https://doi.org/10.3390/nano9050659.

(26) Luongo, G.; Giubileo, F.; Genovese, L.; Iemmo, L.; Martucciello, N.; Di Bartolomeo, A. I-V and C-V Characterization of a High-Responsivity Graphene/Silicon Photodiode with Embedded MOS Capacitor. *Nanomaterials* **2017**, *7* (7), 158. https://doi.org/10.3390/nano7070158.

(27) Di Bartolomeo, A.; Luongo, G.; Giubileo, F.; Funicello, N.; Niu, G.; Schroeder, T.; Lisker, M.; Lupina, G. Hybrid Graphene/Silicon Schottky Photodiode with Intrinsic Gating Effect. *2D Mater.* **2017**, *4* (2), 025075. https://doi.org/10.1088/2053-1583/aa6aa0.

(28) Riazimehr, S.; Kataria, S.; Bornemann, R.; Haring Bolívar, P.; Ruiz, F. J. G.; Engström, O.; Godoy, A.; Lemme, M. C. High Photocurrent in Gated Graphene–Silicon Hybrid Photodiodes. *ACS Photonics* **2017**, *4* (6), 1506–1514. https://doi.org/10.1021/acsphotonics.7b00285.

(29) Luongo, G.; Di Bartolomeo, A.; Giubileo, F.; Chavarin, C. A.; Wenger, C. Electronic Properties of Graphene/p-Silicon Schottky Junction. *J. Phys. D: Appl. Phys.* **2018**, *51* (25), 255305. https://doi.org/10.1088/1361-6463/aac562.

(30) Lupina, G.; Kitzmann, J.; Costina, I.; Lukosius, M.; Wenger, C.; Wolff, A.; Vaziri, S.; Östling, M.; Pasternak, I.; Krajewska, A.; Strupinski, W.; Kataria, S.; Gahoi, A.; Lemme, M. C.; Ruhl, G.; Zoth, G.; Luxenhofer, O.; Mehr, W. Residual Metallic Contamination of Transferred Chemical Vapor Deposited Graphene. *ACS Nano* **2015**, *9* (5), 4776–4785. https://doi.org/10.1021/acsnano.5b01261.

(31) Cheung, S. K.; Cheung, N. W. Extraction of Schottky Diode Parameters from Forward Current-voltage Characteristics. *Appl. Phys. Lett.* **1986**, *49* (2), 85–87. https://doi.org/10.1063/1.97359.

(32) Bartolomeo, A. D.; Giubileo, F.; Luongo, G.; Iemmo, L.; Martucciello, N.; Niu, G.; Fraschke, M.; Skibitzki, O.; Schroeder, T.; Lupina, G. Tunable Schottky Barrier and High Responsivity in Graphene/Si-Nanotip Optoelectronic Device. *2D Mater.* **2016**, *4* (1), 015024. https://doi.org/10.1088/2053-1583/4/1/015024.

(33) Tomer, D.; Rajput, S.; Hudy, L. J.; Li, C. H.; Li, L. Inhomogeneity in Barrier Height at Graphene/Si (GaAs) Schottky Junctions. *Nanotechnology* **2015**, *26* (21), 215702. https://doi.org/10.1088/0957-4484/26/21/215702.

(34) Werner, J. H.; Güttler, H. H. Barrier Inhomogeneities at Schottky Contacts. *Journal of Applied Physics* **1991**, *69* (3), 1522–1533. https://doi.org/10.1063/1.347243.

(35) Parui, S.; Ruiter, R.; Zomer, P. J.; Wojtaszek, M.; van Wees, B. J.; Banerjee, T. Temperature Dependent Transport Characteristics of Graphene/n-Si Diodes. *Journal of Applied Physics* **2014**, *116* (24), 244505. https://doi.org/10.1063/1.4905110.

(36) Shi, E.; Li, H.; Yang, L.; Zhang, L.; Li, Z.; Li, P.; Shang, Y.; Wu, S.; Li, X.; Wei, J.; Wang, K.; Zhu, H.; Wu, D.; Fang, Y.; Cao, A. Colloidal Antireflection Coating Improves Graphene–Silicon Solar Cells. *Nano Lett.* **2013**, *13* (4), 1776–1781. https://doi.org/10.1021/nl400353f.

(37) Yang, H.; Heo, J.; Park, S.; Song, H. J.; Seo, D. H.; Byun, K.-E.; Kim, P.; Yoo, I.; Chung, H.-J.; Kim, K. Graphene Barristor, a Triode Device with a Gate-Controlled Schottky Barrier. *Science* **2012**, *336* (6085), 1140–1143. https://doi.org/10.1126/science.1220527.

(38) Yim, C.; McEvoy, N.; Duesberg, G. S. Characterization of Graphene-Silicon Schottky Barrier Diodes Using Impedance Spectroscopy. *Appl. Phys. Lett.* **2013**, *103* (19), 193106. https://doi.org/10.1063/1.4829140.

(39) Courtin, J.; Le Gall, S.; Chrétien, P.; Moréac, A.; Delhaye, G.; Lépine, B.; Tricot, S.; Turban, P.; Schieffer, P.; Le Breton, J.-C. A Low Schottky Barrier Height and Transport Mechanism in Gold–Graphene–Silicon (001) Heterojunctions. *Nanoscale Adv.* **2019**, *1* (9), 3372–3378. https://doi.org/10.1039/C9NA00393B.

(40) Singh, A.; Uddin, Md. A.; Sudarshan, T.; Koley, G. Tunable Reverse-Biased Graphene/Silicon Heterojunction Schottky Diode Sensor. *Small* **2014**, *10* (8), 1555–1565. https://doi.org/10.1002/smll.201302818.

(41) Kim, H.-Y.; Lee, K.; McEvoy, N.; Yim, C.; Duesberg, G. S. Chemically Modulated Graphene Diodes. *Nano Lett.* **2013**, *13* (5), 2182–2188. https://doi.org/10.1021/nl400674k.

(42) Sinha, D.; Lee, J. U. Ideal Graphene/Silicon Schottky Junction Diodes. *Nano Lett.* **2014**, *14* (8), 4660–4664. https://doi.org/10.1021/nl501735k.

(43) Javadi, M.; Noroozi, A.; Abdi, Y. Kinetics of Charge Carriers across a Graphene-Silicon Schottky Junction. *Phys. Rev. Applied* **2020**, *14* (6), 064048. https://doi.org/10.1103/PhysRevApplied.14.064048.

(44) Wang, C.; Dong, Y.; Lu, Z.; Chen, S.; Xu, K.; Ma, Y.; Xu, G.; Zhao, X.; Yu, Y. High Responsivity and High-Speed 1.55 Mm Infrared Photodetector from Self-Powered Graphene/Si Heterojunction. *Sensors and Actuators A: Physical* **2019**, *291*, 87–92. https://doi.org/10.1016/j.sna.2019.03.054.

(45) Tang, Y.; Chen, J. High Responsivity of Gr/ n-Si Schottky Junction near-Infrared Photodetector. *Superlattices and Microstructures* **2021**, *150*, 106803. https://doi.org/10.1016/j.spmi.2021.106803.

(46) High Efficiency Graphene Solar Cells by Chemical Doping | Nano Letters https://pubs.acs.org/doi/pdf/10.1021/nl204414u (accessed May 5, 2021).

(47) Feng, S.; Dong, B.; Lu, Y.; Yin, L.; Wei, B.; Wang, J.; Lin, S. Graphene/p-AlGaN/p-GaN Electron Tunnelling Light Emitting Diodes with High External Quantum Efficiency. *Nano Energy* **2019**, *60*, 836–840. https://doi.org/10.1016/j.nanoen.2019.04.007.

(48) Ruan, K.; Ding, K.; Wang, Y.; Diao, S.; Shao, Z.; Zhang, X.; Jie, J. Flexible Graphene/Silicon Heterojunction Solar Cells. *J. Mater. Chem. A* **2015**, *3* (27), 14370–14377. https://doi.org/10.1039/C5TA03652F.

(49) Jensen, K. L. Electron Emission Theory and Its Application: Fowler–Nordheim Equation and Beyond. *J. Vac. Sci. Technol. B* **2003**, *21* (4), 1528. https://doi.org/10.1116/1.1573664.

(50) Di Bartolomeo, A.; Yang, Y.; Rinzan, M. B. M.; Boyd, A. K.; Barbara, P. Record Endurance for Single-Walled Carbon Nanotube–Based Memory Cell. *Nanoscale Research Letters* **2010**, *5* (11), 1852. https://doi.org/10.1007/s11671-010-9727-6.